\documentclass[twocolumn,american,prl,amsmath,amssymb,superscriptaddress,showpacs]{revtex4}
\usepackage[T1]{fontenc}
\usepackage[latin1]{inputenc}
\usepackage{amsmath}
\usepackage{graphicx}
\usepackage{amssymb}
\usepackage{esint}

\makeatletter
\@ifundefined{textcolor}{}
{%
 \definecolor{BLACK}{gray}{0}
 \definecolor{WHITE}{gray}{1}
 \definecolor{RED}{rgb}{1,0,0}
 \definecolor{GREEN}{rgb}{0,1,0}
 \definecolor{BLUE}{rgb}{0,0,1}
 \definecolor{CYAN}{cmyk}{1,0,0,0}
 \definecolor{MAGENTA}{cmyk}{0,1,0,0}
 \definecolor{YELLOW}{cmyk}{0,0,1,0}
 }

\@ifundefined{definecolor}
 {\usepackage{color}}{}

\makeatletter
\@ifundefined{textcolor}{}{\definecolor{BLACK}{gray}{0}\definecolor{WHITE}{gray}{1}\definecolor{RED}{rgb}{1,0,0}\definecolor{GREEN}{rgb}{0,1,0}\definecolor{BLUE}{rgb}{0,0,1}\definecolor{CYAN}{cmyk}{1,0,0,0}\definecolor{MAGENTA}{cmyk}{0,1,0,0}\definecolor{YELLOW}{cmyk}{0,0,1,0}}

\@ifundefined{definecolor}{\@ifundefined{definecolor}
 {\usepackage{color}}{}
}{}
\usepackage{dcolumn}
\usepackage{bm}
\usepackage{rotating}\usepackage{psfrag}

\def\kbar{{\mathchar'26\mkern-9mu k}}
\makeatother

\usepackage{babel}

\makeatother

\begin{document}

\title{Between a metal and an insulator: the critical state of the Anderson
transition}

\author{Gabriel Lemari\'e}

\altaffiliation{Present address: Service de Physique de l'Etat Condens\'e (CNRS URA 2464), IRAMIS/SPEC, CEA Saclay, F-91191 Gif-sur-Yvette, France}

\affiliation{Laboratoire Kastler Brossel, UPMC-Paris 6, ENS, CNRS; 4 Place Jussieu,
F-75005 Paris, France}

\author{Hans Lignier}

\altaffiliation{Present address: Laboratoire Aim\'e Cotton, Universit\'e Paris-Sud, Bat. 505, Campus d'Orsay, F-91405 Orsay Cedex, France}

\affiliation{Laboratoire de Physique des Lasers, Atomes et Mol\'ecules, Universit\'e
Lille 1 Sciences et Technologies, UMR CNRS 8523; F-59655 Villeneuve
d'Ascq Cedex, France}

\homepage{www.phlam.univ-lille1.fr/atfr/cq}

\author{Dominique Delande}

\affiliation{Laboratoire Kastler Brossel, UPMC-Paris 6, ENS, CNRS; 4 Place Jussieu,
F-75005 Paris, France}

\author{Pascal Szriftgiser}

\author{Jean-Claude Garreau }

\affiliation{Laboratoire de Physique des Lasers, Atomes et Mol\'ecules, Universit\'e
Lille 1 Sciences et Technologies, UMR CNRS 8523; F-59655 Villeneuve
d'Ascq Cedex, France}

\homepage{www.phlam.univ-lille1.fr/atfr/cq}

\date{\today}
\begin{abstract}
Using a three-frequency one-dimensional kicked rotor experimentally realized
with a cold atomic gas,
we study the transport properties at the critical point of the metal-insulator
Anderson transition. We accurately measure the time-evolution of an initially
localized wavepacket and show that it displays at
the critical point a scaling invariance characteristic of this second-order
phase transition. The shape of the momentum distribution at the critical point
is found to be in excellent agreement with the analytical form deduced from self-consistent theory of localization. 
\end{abstract}
\maketitle
Different phase transitions observed in various fields of physics
often share similar characteristics~\cite{Cardy:book96}. Of special
interest is the behavior of the system at the critical point (for
example scale invariance) and in its immediate vicinity (e.g. divergence
of a characteristic length scale). The advent of cold atom physics
has offered new possibilities of direct experimental observation of
such characteristics of quantum phase transitions. In this letter,
we show that the Anderson metal-insulator transition (which
has only recently been observed with atomic matter waves~\cite{Chabe:PRL08})
obeys scale invariance at the threshold, defining a new state of matter
between a metal and an insulator.

The Anderson transition takes place in 3-dimensional (3D) disordered
non-interacting systems in the mesoscopic regime (where the transport
is coherent). It involves a metallic phase at low disorder associated
with an essentially diffusive transport, and an insulating phase at
large disorder where transport over long distance is inhibited by
interference effects: this is the so-called \emph{Anderson localization}
phenomenon~\cite{Anderson:PR58}. The Anderson transition is a second-order
(continuous) phase transition: On the insulating side, the localization
characteristic length $\ell$ diverges algebraically, $\ell\propto|K-K_{c}|^{-\nu}$
when $K$, the control parameter, approaches the threshold $K_{c}$
of the transition. On the metallic side, similarly, the diffusion
constant vanishes algebraically $D\propto|K-K_{c}|^{s}$. The critical
exponents $s$ and $\nu$ are equal in 3D, and universal (they do
not depend on the microscopic details of the system)~\cite{Wegner:ZFP76}.
Only recently have these theoretical predictions been confirmed experimentally
and the value of $\nu=s$ unambiguously determined~\cite{Chabe:PRL08,Lemarie:PRA09,Lemarie:EPL09}:
$\nu=1.4\pm0.3$ is found perfectly compatible with $\nu=1.57\pm0.02$
obtained from numerical simulations of the 3D Anderson model~\cite{Slevin:PRL99,Slevin:2010}.

%
\begin{figure*}
\begin{centering}
\includegraphics[width=15cm]{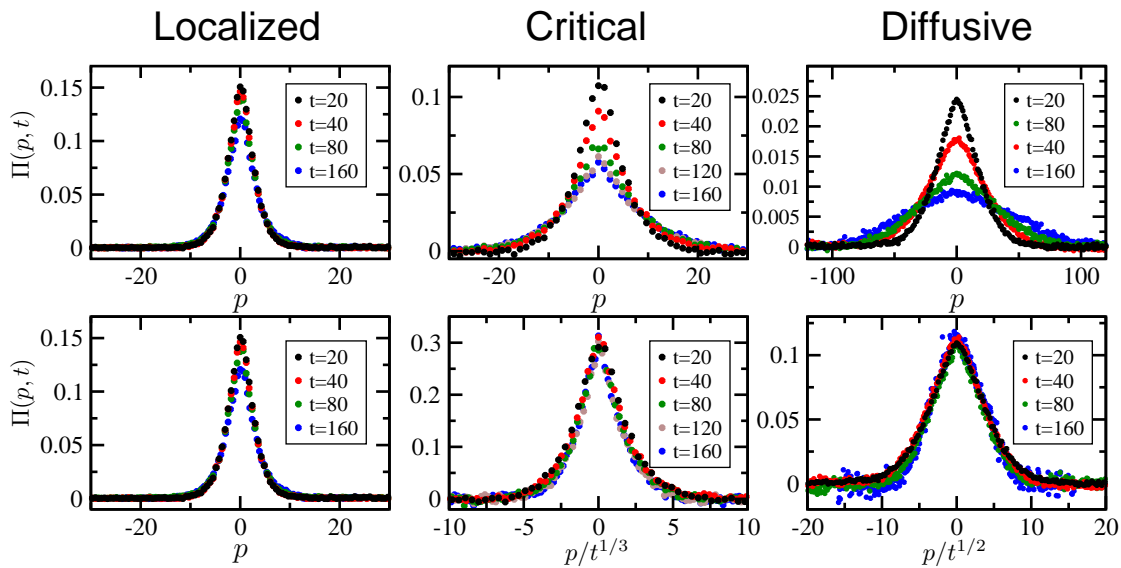} 
\par\end{centering}

\caption{\label{fig:fonc-rescale} First row: Measured AIGF of the quasiperiodic
atomic kicked rotor at different times $t$ in the (left to right)
localized $(K,\varepsilon)=(6,0.1)$, critical $(K=K_c,\varepsilon)=(8,0.38)$
and diffusive $(K,\varepsilon)=(11,0.8)$ regimes. Second row: Appropriate
rescalings of the momentum by $t^{0}$ (localized), $t^{1/3}$ (critical)
or $t^{1/2}$ (diffusive), bring the curves at different times into
coincidence (the vertical scales are also rescaled in order to preserve
normalization). The shapes are different in the three regimes: exponential
localization, Eq.~(\ref{eq:localized}), in the localized regime,
Gaussian shape, Eq.~(\ref{eq:diffusive}), in the diffusive regime,
and the new ``Airy shape'', Eq.~(\ref{eq:fonccritAi}), at the
critical point. Parameters are: $\kbar=2.89$, $\omega_{2}=2\pi\sqrt{7}$,
$\omega_{3}=2\pi\sqrt{17}$. Time is measured in number of kicks and
momentum in units of $2\hbar k_{L}.$ }

\end{figure*}

The state of a disordered system is, in this context, characterized
by its transport properties. One can consider the behavior at large
distances and long times of the (disorder) averaged intensity Green
function (AIGF) which determines the probability $P(\boldsymbol{r},\boldsymbol{r}';t)$
for a particle to go from $\boldsymbol{r}$ to $\boldsymbol{r}'$
in time $t$~\cite{Vollhardt:book91}. In the insulating phase, the
AIGF is a stationary function exponentially localized: 
\begin{equation}
P(\boldsymbol{r},\boldsymbol{r}';t)\sim\exp(-\vert\boldsymbol{r}-\boldsymbol{r}'\vert/2\ell),\ \ \ \ \ \ \ \ \ \mathrm{[localized]}
\label{eq:localized}
\end{equation}
 while in the metallic regime, it is a Gaussian expanding diffusively:
\begin{equation}
P(\boldsymbol{r},\boldsymbol{r}';t)\sim\exp[-(\boldsymbol{r}-\boldsymbol{r}')^{2}/2Dt].\ \ \ \ \ \ \ \ \mathrm{[diffusive]}
\label{eq:diffusive}
\end{equation}
 These two behaviors are however \textit{long} time asymptotics. Indeed,
a localized AIGF is observed only for times $t\gg t_{\ell}$, where
$t_{\ell}$ is the localization time (the time-scale associated to
localization). At the transition, $t_{\ell}\sim\ell^{3}$ diverges
and the system becomes scale invariant. What is the AIGF behavior
at the critical point? In the following, we show that it scales as:
\begin{equation}
P(\boldsymbol{r},\boldsymbol{r}';t)\sim\exp\left[-\alpha\vert\boldsymbol{r}-\boldsymbol{r}'\vert^{3/2}/t^{1/2}\right]\;,\ \ \ \ \ \ \ \mathrm{[critical]}
\label{eq:asymptAIGF}
\end{equation}
 where $\alpha$ is a known (measurable) quantity. This defines a
new state, since the shape does not change with time. Such a state
of matter, intermediate between an insulator and a metal at all scales,
has never been directly observed experimentally, although interesting
results have recently been published for ultrasound waves in the localized
regime~\cite{Faez:PRL09}. The purpose of this letter is to report
the first experimental characterization of such a critical state of
the Anderson transition.

In cold atomic gases, it is possible to prepare the system in a localized
state and follow its evolution over time; this constitutes an experimental
measurement of the (A)IGF~\cite{Raizen:PRL95,Bouyer:N08,Chabe:PRL08},
which is impossible to achieve in state of art solid state
physics. Observing the 3D Anderson transition in configuration space
with cold atoms requires a disordered potential with a correlation
length comparable to the de Broglie wavelength~\cite{Kuhn:2007}
which has not yet been achieved. We have recently shown~\cite{Chabe:PRL08,Lemarie:PRA09}
that it is nevertheless possible to observe Anderson localization
and the Anderson transition in \emph{momentum space} by using a different
system, the atomic kicked rotor (described below), where the chaotic
nature of the classical motion replaces the disordered potential.

Our atom-optics system (see \cite{Lemarie:PRA09} for a detailed description)
consists in a cloud of laser-cooled cesium atoms (FWHM of the momentum
distribution of 8$\hbar k_{L}$) interacting with a pulsed (period
$T_{1}=27.778$ $\mu$s), far detuned ($\Delta=-11.3$ GHz) standing
wave (wavenumber $k_{L}=7.4\times10^{6}$ m$^{-1}$ and one way intensity
$I_{0}=150$ mW). The amplitude of the kicks is modulated with two
frequencies $\omega_{2}$ and $\omega_{3}$. The Hamiltonian reads:
\begin{equation}
H=\frac{p^{2}}{2}+K\cos x\left[1+\varepsilon\cos\left(\omega_{2}t\right)\cos\left(\omega_{3}t\right)\right]\sum_{n=0}^{N-1}\delta(t-n)\;,
\label{eq:H}
\end{equation}
 where time is measured in units of $T_{1}$, space in units of $(2k_{L})^{-1}$,
momentum in units of $2\hbar k_{L}/\kbar$, with $\kbar=4\hbar k_{L}^{2}T_{1}/M=2.89$
($M$ is the atom mass) playing the role of an effective Planck constant
($[x,p]=i\kbar$) and $K$ is the average kick amplitude. The kicks
are short enough (duration $\tau=0.8$ $\mu$s) as compared to the
atom dynamics so that they can be considered as Dirac delta functions.
Decoherence processes, analyzed in detail in \cite{Lemarie:PRA09}
are negligible for the typical duration of the experiment $t\simeq160$
kicks.

If $\omega_{2}$, $\omega_{3},$ $\pi$ and $\kbar$ are incommensurate,
this 1D quasiperiodic kicked rotor has been shown to be equivalent
to a 3D disordered anisotropic system~\cite{Casati:PRL89,Basko:PRL03,Lemarie:EPL09}
and to display an Anderson metal-insulator transition, as evidenced
by the fact that it belongs to the universality class of the 3D Anderson
model \cite{Lemarie:EPL09,Slevin:PRL99}, i.e. has the same critical
exponent $\nu$. Here, the localization manifests itself in momentum
space instead of configuration space. We thus expect the AIGF to take
simpler forms in momentum space, with expressions similar to Eqs.~(\ref{eq:localized})-(\ref{eq:asymptAIGF})
(simply replacing position ${\boldsymbol{r}}$ by momentum ${p}$).
In order to avoid confusion, we will use the notation $\Pi({p},{p}';t)$
for the AIGF in momentum space. Thus, an initial momentum distribution
$W({p},t=0)$ will on average evolve at time $t$ to: 
\begin{equation}
W({p},t)=\int{\Pi({p},{p}';t)\ W({p}',0)\ \mathrm{d}{p}'}\;.\label{W.eq}
\end{equation}

Experimentally, we are able to measure the momentum distribution at
the end of a pulse sequence (up to 160 kicks), using Raman stimulated
transitions (see \cite{AP:DiodeMod:EPJD99,AP:RamanSpectro:PRA01}
for details). The initial state $W({p},t=0)$ is a thermal momentum
distribution whose width is much smaller than the width of the final
distribution, and can thus be approximated by a $\delta$-function
$\delta({p})$ in Eq.~(\ref{W.eq}). As a consequence, the final
momentum distribution $W({p},t)$ faithfully measures the intensity
Green function $\Pi({p},t)\equiv\Pi({p},0;t).$

%
\begin{figure}
\includegraphics[width=6.2cm,angle=90]{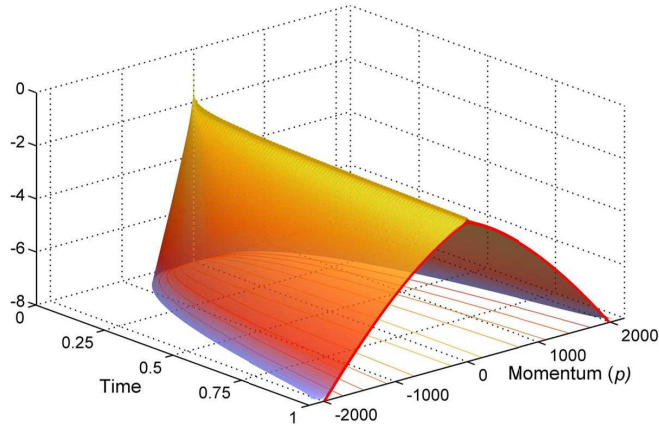} \centering{} \caption{\label{fig:fonc-crit-3d} Numerically simulated time-evolution of an initially localized momentum distribution (log scale),
for the quasiperiodic kicked rotor at the critical point of the Anderson transition.
The spreading follows an anomalous diffusion, with $\langle p^{2}(t)\rangle\propto t^{2/3},$
and the shape is preserved (i.e. scale invariant), being neither exponentially
(as it is in the localized regime), nor Gaussian (as it is in the
diffusive regime). The analytic prediction, Eq.~(\ref{eq:fonccritAi})
is shown as the red thick curve at 1 million kicks. The agreement
is excellent, without any adjustable parameter. Parameters are those
of fig.~\ref{fig:fonc-rescale}. Time is measured in millions of
kicks and momentum in units of two recoil momenta $2\hbar k_{L}.$ }

\end{figure}

Figure~\ref{fig:fonc-rescale} shows the experimentally measured
AIGF $\Pi(p,t)$ at various times in the localized, critical and diffusive
regimes. The three regimes obey different scaling laws. In the localized
regime (left column), the momentum distribution is localized -- i.e.
it is time-independent -- at long times and thus scales as $t^{0}.$
In the diffusive regime, the average kinetic energy $\langle p^{2}(t)\rangle$
increases linearly with time, so that the typical momentum scales
as $t^{1/2};$ this is manifest in the broadening of the distribution
with time seen in the right column. At the critical point of the Anderson
transition, we observe \cite{Chabe:PRL08,Lemarie:PRA09}, as predicted
by the one parameter scaling theory \cite{Abrahams:PRL79,Ohtsuki:JPSJ97},
an anomalous diffusion $\langle p^{2}\rangle(t)\sim t^{2/3}$. This
implies that the typical momentum scales as $t^{1/3}$ leading to
a slower broadening of the distribution (middle column). If the raw
experimental data are rescaled according to these laws (lower row
in Fig.~\ref{fig:fonc-rescale}), i.e. plotted vs. $p t^{0}$,
$p t^{-1/3}$ and $p t^{-1/2}$ in the localized, critical and diffusive
regimes respectively, curves taken at various times coincide, which
constitutes an experimental \emph{proof} of the validity of the scaling
laws 
\footnote{Extracting the precise location of the critical point is not straightforward,
especially because the experimental data are limited to 160 kicks.
As explained in~\cite{Lemarie:PRA09}, finite-time-scaling makes
it possible to determine it with a reasonably small uncertainty, of
the order of 0.3 on the value of $K.$ The experimentally observed
value agrees very well with the one extracted from numerical simulations
without any adjustable parameter.}. 
The shapes of the distributions are
different in the various regimes: exponential shape in the localized
regime, Gaussian shape in the diffusive regime. The intermediate shape
at the critical point is discussed below.

Figure~\ref{fig:fonc-rescale} is a clear manifestation of the scale
invariance at the critical point. The anomalous diffusion is not a
transient behavior and the AIGF keeps the same shape at the
critical point. However, slightly off the critical point, the AIGF
tends gradually to either a localized or diffusive behavior, following
the anomalous diffusion only for short times. To confirm this observation
of scale invariance over a time scale larger than $160$ kicks, we
performed numerical simulations of the critical dynamics up to $t=10^{6}$
kicks. The result is shown in Fig.~\ref{fig:fonc-crit-3d}. The
advantage of numerical simulations is that it is possible to explore
the tails of the momentum distributions (hidden by noise in a real
experiment). The anomalous diffusion -- with the characteristic sub-diffusive
$t^{1/3}$ scaling -- is clearly visible. Obviously, the distribution
is neither exponentially shaped (which would result in straight lines
in the logarithmic plot), nor has a Gaussian shape (a parabola in
the logarithmic plot).

%
\begin{figure}
\centering{}\includegraphics[width=8cm]{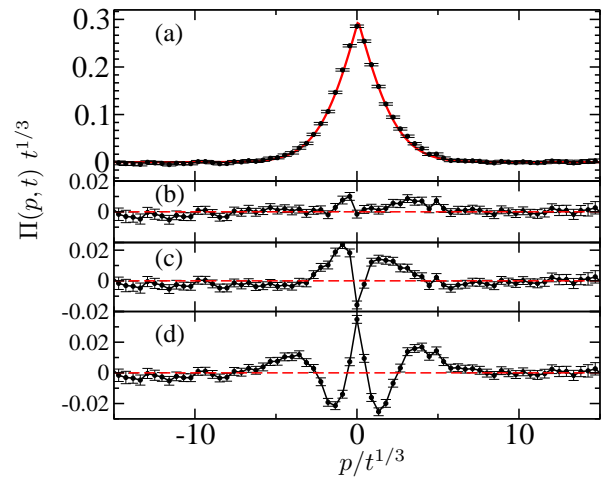} \caption{\label{fig:fonc-crit-fit} (color online) (a) Experimental data for
the rescaled critical AIGF (see fig. \ref{fig:fonc-rescale}) averaged
over time (black circles with error bars) and a fit given by Eq.~(\ref{eq:fonccritAi}),
with $\rho$ the only fitting parameter. The agreement is clearly
excellent. The fitted value $\rho$ is found compatible with $\rho=\Gamma(2/3)\Lambda_{c}/3$.
The residual does not significantly differ from zero [panel (b)].
Fits by an exponentially localized (c) or a Gaussian (d) distribution
show significant deviations.}

\end{figure}

The form of the critical AIGF can be deduced from the self-consistent
theory of localization \cite{Vollhardt:book91}. This mean-field theory
describes quantum transport in disordered systems at large distances
and for long times \cite{bart:NP08}. It has been shown to be relevant
for the 1D periodically kicked rotor \cite{Atland:PRL93} and correctly
predicts a metal-insulator transition in three dimensions and the
anomalous diffusion at the threshold: $D(\omega)\sim(-i\omega)^{1/3}$
with $\omega$ the frequency conjugated to time \cite{Shapiro:PRB82}.
(the 1/3 exponent is the counterpart in the frequency domain of the
anomalous diffusion $\langle p^{2}\rangle(t)\sim t^{2/3}$ in the time domain). Using this critical behavior, we can compute the AIGF for the quasiperiodic kicked rotor~\cite{Lemarie:these}. The details of the calculation will be published elsewhere; we obtain: \begin{equation}
\Pi(p,t)=\frac{3}{2}\left(3\rho^{3/2}t\right)^{-1/3}\text{Ai}\left[\left(3\rho^{3/2}t\right)^{-1/3}\vert p\vert\right]\;,\label{eq:fonccritAi}\end{equation}
 where $\rho$ is a parameter directly related to the critical quantity
$\Lambda_{c}=\underset{t\rightarrow\infty}{\lim}\langle p^{2}\rangle/t^{2/3}$
(see \cite{Chabe:PRL08,Lemarie:PRA09}) via $\rho=\Gamma(2/3)\Lambda_{c}/3$,
where $\Gamma$ is the Gamma function and $\text{Ai}(x)$ is the Airy function. This expression is used for
the plot in Fig.~\ref{fig:fonc-crit-3d}. The asymptotic form Eq.~(\ref{eq:asymptAIGF})
comes simply from the limiting behavior of the Airy function
for large $x$ and is found perfectly intermediate between the exponential
(localized) and the Gaussian (diffusive) shapes.

The analytic prediction, Eq.~(\ref{eq:fonccritAi}), matches very
well the shape obtained from numerical simulations of the quasiperiodically
kicked rotor shown in Fig.~\ref{fig:fonc-crit-3d}. The only noticeable
difference is near $p=0,$ where the result of the numerical simulation
is slightly larger than the analytic prediction. Note that this is
only observed at very long times, beyond 1000 kicks; this phenomenon
is currently under study. On the time scale of the experiment (160
kicks), this effect is invisible.

Figure~\ref{fig:fonc-crit-fit} shows the comparison between the
experimentally measured critical AIGF and the analytic prediction, Eq.~(\ref{eq:fonccritAi}).
The only fitting parameter is the global scale $\rho$ (found in excellent
agreement with $\rho=\Gamma(2/3)\Lambda_{c}/3$). Although it is visually
not obvious to distinguish the observed shape from either an exponential
shape or a Gaussian shape, a careful fitting procedure gives a clear
cut result. The residual between the observed distribution and the
analytic prediction (\ref{eq:fonccritAi}), shown in panel (b), is
consistently zero (within the error bars), while a fit with an exponential
shape, panel (c), of a Gaussian shape, panel (d), displays significant
deviations. This is fully confirmed by a quantitative check of the
quality of the fit. The fit by the Airy function gives a $\chi^{2}$
per degree of freedom equal to 1.09 -- i.e. perfectly acceptable --
while the exponential fit gives 4.5 per degree of freedom and the
Gaussian fit 8.8, two unacceptably large values. This clearly shows
that the self-consistent theory of localization accounts for the critical
AIGF and its scale invariance.

In conclusion, we have studied experimentally the transport at the
threshold of the Anderson transition. It obeys scale invariance, one
fundamental property of this second-order phase transition, and this
defines a new state, between an insulator and a metal. Its analytic
form can be deduced from the self-consistent theory of localization.
Work is in progress to allow experimental observations at longer times,
which should allow us to characterize the small deviations observed
numerically, whose origin could be multifractality. 
\begin{acknowledgments}
The authors acknowledge Narei Mart{\'{i}}nez for her help with
the experiment. This work was partially financed by  
Ministry of Higher Education and Research,
Nord-Pas de Calais Regional Council and FEDER through the ``Contrat de
Projets Etat Region (CPER) 2007-2013'' and was granted access to the HPC resources
of IDRIS under the allocation 2009-96089 made by GENCI (Grand Equipement
National de Calcul Intensif). 
\end{acknowledgments}


\begin{thebibliography}{22}
\expandafter\ifx\csname natexlab\endcsname\relax\def\natexlab#1{#1}\fi
\expandafter\ifx\csname bibnamefont\endcsname\relax
  \def\bibnamefont#1{#1}\fi
\expandafter\ifx\csname bibfnamefont\endcsname\relax
  \def\bibfnamefont#1{#1}\fi
\expandafter\ifx\csname citenamefont\endcsname\relax
  \def\citenamefont#1{#1}\fi
\expandafter\ifx\csname url\endcsname\relax
  \def\url#1{\texttt{#1}}\fi
\expandafter\ifx\csname urlprefix\endcsname\relax\def\urlprefix{URL }\fi
\providecommand{\bibinfo}[2]{#2}
\providecommand{\eprint}[2][]{\url{#2}}

\bibitem[{\citenamefont{Cardy}(1996)}]{Cardy:book96}
\bibinfo{author}{\bibfnamefont{J.}~\bibnamefont{Cardy}},
  \emph{\bibinfo{title}{Scaling and Renormalization in Statistical Physics}}
  (\bibinfo{publisher}{Cambridge University Press},
  \bibinfo{address}{Cambridge}, \bibinfo{year}{1996}).

\bibitem[{\citenamefont{Chab{\'e} et~al.}(2008)\citenamefont{Chab{\'e},
  Lemari{\'e}, Gr{\'e}maud, Delande, Szriftgiser, and Garreau}}]{Chabe:PRL08}
\bibinfo{author}{\bibfnamefont{J.}~\bibnamefont{Chab{\'e}}},
  \bibinfo{author}{\bibfnamefont{G.}~\bibnamefont{Lemari{\'e}}},
  \bibinfo{author}{\bibfnamefont{B.}~\bibnamefont{Gr{\'e}maud}},
  \bibinfo{author}{\bibfnamefont{D.}~\bibnamefont{Delande}},
  \bibinfo{author}{\bibfnamefont{P.}~\bibnamefont{Szriftgiser}},
  \bibnamefont{and} \bibinfo{author}{\bibfnamefont{J.~C.}
  \bibnamefont{Garreau}}, \bibinfo{journal}{Phys. Rev. Lett.}
  \textbf{\bibinfo{volume}{101}}, \bibinfo{pages}{255702}
  (\bibinfo{year}{2008}).

\bibitem[{\citenamefont{Anderson}(1958)}]{Anderson:PR58}
\bibinfo{author}{\bibfnamefont{P.~W.} \bibnamefont{Anderson}},
  \bibinfo{journal}{Phys. Rev.} \textbf{\bibinfo{volume}{109}},
  \bibinfo{pages}{1492} (\bibinfo{year}{1958}).

\bibitem[{\citenamefont{Wegner}(1976)}]{Wegner:ZFP76}
\bibinfo{author}{\bibfnamefont{F.}~\bibnamefont{Wegner}}, \bibinfo{journal}{Z.
  Phys.} \textbf{\bibinfo{volume}{B25}}, \bibinfo{pages}{327}
  (\bibinfo{year}{1976}).

\bibitem[{\citenamefont{Lemari\'e
  et~al.}(2009{\natexlab{a}})\citenamefont{Lemari\'e, Chab\'e, Szriftgiser,
  Garreau, Gr\'emaud, and Delande}}]{Lemarie:PRA09}
\bibinfo{author}{\bibfnamefont{G.}~\bibnamefont{Lemari\'e}},
  \bibinfo{author}{\bibfnamefont{J.}~\bibnamefont{Chab\'e}},
  \bibinfo{author}{\bibfnamefont{P.}~\bibnamefont{Szriftgiser}},
  \bibinfo{author}{\bibfnamefont{J.~C.} \bibnamefont{Garreau}},
  \bibinfo{author}{\bibfnamefont{B.}~\bibnamefont{Gr\'emaud}},
  \bibnamefont{and} \bibinfo{author}{\bibfnamefont{D.}~\bibnamefont{Delande}},
  \bibinfo{journal}{Phys. Rev. A} \textbf{\bibinfo{volume}{80}},
  \bibinfo{pages}{043626} (\bibinfo{year}{2009}{\natexlab{a}}).

\bibitem[{\citenamefont{Lemari\'e
  et~al.}(2009{\natexlab{b}})\citenamefont{Lemari\'e, Gr\'emaud, and
  Delande}}]{Lemarie:EPL09}
\bibinfo{author}{\bibfnamefont{G.}~\bibnamefont{Lemari\'e}},
  \bibinfo{author}{\bibfnamefont{B.}~\bibnamefont{Gr\'emaud}},
  \bibnamefont{and} \bibinfo{author}{\bibfnamefont{D.}~\bibnamefont{Delande}},
  \bibinfo{journal}{Europhys. Lett.} \textbf{\bibinfo{volume}{87}}, \bibinfo{pages}{37007}
  (\bibinfo{year}{2009}{\natexlab{b}}).

\bibitem[{\citenamefont{Slevin and Ohtsuki}(1999)}]{Slevin:PRL99}
\bibinfo{author}{\bibfnamefont{K.}~\bibnamefont{Slevin}} \bibnamefont{and}
  \bibinfo{author}{\bibfnamefont{T.}~\bibnamefont{Ohtsuki}},
  \bibinfo{journal}{Phys. Rev. Lett.} \textbf{\bibinfo{volume}{82}},
  \bibinfo{pages}{382} (\bibinfo{year}{1999}).

\bibitem{Slevin:2010} A. Rodriguez et al., arXiv:1005.0515.

\bibitem[{\citenamefont{Vollhardt and W\"olfle}(1992)}]{Vollhardt:book91}
\bibinfo{author}{\bibfnamefont{D.}~\bibnamefont{Vollhardt}} \bibnamefont{and}
  \bibinfo{author}{\bibfnamefont{P.}~\bibnamefont{W\"olfle}}, in
  \emph{\bibinfo{booktitle}{Electronic Phase Transitions}}, edited by
  \bibinfo{editor}{\bibfnamefont{W.}~\bibnamefont{Hanke}} \bibnamefont{and}
  \bibinfo{editor}{\bibfnamefont{Y.~V.} \bibnamefont{Kopaev}}
  (\bibinfo{publisher}{Elsevier}, \bibinfo{address}{Amsterdam},
  \bibinfo{year}{1992}), pp. \bibinfo{pages}{1--78}.

\bibitem[{\citenamefont{Faez et~al.}(2009)\citenamefont{Faez, Strybulevych,
  Page, Lagendijk, and van Tiggelen}}]{Faez:PRL09}
\bibinfo{author}{\bibfnamefont{S.}~\bibnamefont{Faez}},
  \bibinfo{author}{\bibfnamefont{A.}~\bibnamefont{Strybulevych}},
  \bibinfo{author}{\bibfnamefont{J.~H.} \bibnamefont{Page}},
  \bibinfo{author}{\bibfnamefont{A.}~\bibnamefont{Lagendijk}},
  \bibnamefont{and} \bibinfo{author}{\bibfnamefont{B.~A.} \bibnamefont{van
  Tiggelen}}, \bibinfo{journal}{Phys. Rev. Lett.}
  \textbf{\bibinfo{volume}{103}}, \bibinfo{pages}{155703}
  (\bibinfo{year}{2009}).

\bibitem[{\citenamefont{Moore et~al.}(1995)\citenamefont{Moore, Robinson,
  Bharucha, Sundaram, and Raizen}}]{Raizen:PRL95}
\bibinfo{author}{\bibfnamefont{F.~L.} \bibnamefont{Moore}},
  \bibinfo{author}{\bibfnamefont{J.~C.} \bibnamefont{Robinson}},
  \bibinfo{author}{\bibfnamefont{C.~F.} \bibnamefont{Bharucha}},
  \bibinfo{author}{\bibfnamefont{B.}~\bibnamefont{Sundaram}}, \bibnamefont{and}
  \bibinfo{author}{\bibfnamefont{M.~G.} \bibnamefont{Raizen}},
  \bibinfo{journal}{Phys. Rev. Lett.} \textbf{\bibinfo{volume}{75}},
  \bibinfo{pages}{4598} (\bibinfo{year}{1995}).

\bibitem[{\citenamefont{Billy et~al.}(2008)\citenamefont{Billy, Josse, Zuo,
  Bernard, Hambrecht, Lugan, Cl{\'e}ment, Sanchez-Palencia, Bouyer, and
  Aspect}}]{Bouyer:N08}
\bibinfo{author}{\bibfnamefont{J.}~\bibnamefont{Billy}},
  \bibinfo{author}{\bibfnamefont{V.}~\bibnamefont{Josse}},
  \bibinfo{author}{\bibfnamefont{Z.}~\bibnamefont{Zuo}},
  \bibinfo{author}{\bibfnamefont{A.}~\bibnamefont{Bernard}},
  \bibinfo{author}{\bibfnamefont{B.}~\bibnamefont{Hambrecht}},
  \bibinfo{author}{\bibfnamefont{P.}~\bibnamefont{Lugan}},
  \bibinfo{author}{\bibfnamefont{D.}~\bibnamefont{Cl{\'e}ment}},
  \bibinfo{author}{\bibfnamefont{L.}~\bibnamefont{Sanchez-Palencia}},
  \bibinfo{author}{\bibfnamefont{P.}~\bibnamefont{Bouyer}}, \bibnamefont{and}
  \bibinfo{author}{\bibfnamefont{A.}~\bibnamefont{Aspect}},
  \bibinfo{journal}{Nature} \textbf{\bibinfo{volume}{453}},
  \bibinfo{pages}{891} (\bibinfo{year}{2008}).

\bibitem[{\citenamefont{Kuhn et~al.}(2007)\citenamefont{Kuhn, Sigwarth,
  Miniatura, Delande, and M\"uller}}]{Kuhn:2007}
\bibinfo{author}{\bibfnamefont{R.}~\bibnamefont{Kuhn}},
  \bibinfo{author}{\bibfnamefont{O.}~\bibnamefont{Sigwarth}},
  \bibinfo{author}{\bibfnamefont{C.}~\bibnamefont{Miniatura}},
  \bibinfo{author}{\bibfnamefont{D.}~\bibnamefont{Delande}}, \bibnamefont{and}
  \bibinfo{author}{\bibfnamefont{C.}~\bibnamefont{M\"uller}},
  \bibinfo{journal}{New J. Phys.} \textbf{\bibinfo{volume}{9}},
  \bibinfo{pages}{161} (\bibinfo{year}{2007}).

\bibitem[{\citenamefont{Casati et~al.}(1989)\citenamefont{Casati, Guarneri, and
  Shepelyansky}}]{Casati:PRL89}
\bibinfo{author}{\bibfnamefont{G.}~\bibnamefont{Casati}},
  \bibinfo{author}{\bibfnamefont{I.}~\bibnamefont{Guarneri}}, \bibnamefont{and}
  \bibinfo{author}{\bibfnamefont{D.~L.} \bibnamefont{Shepelyansky}},
  \bibinfo{journal}{Phys. Rev. Lett.} \textbf{\bibinfo{volume}{62}},
  \bibinfo{pages}{345} (\bibinfo{year}{1989}).

\bibitem[{\citenamefont{Basko et~al.}(2003)\citenamefont{Basko, Skvortsov, and
  Kravtsov}}]{Basko:PRL03}
\bibinfo{author}{\bibfnamefont{D.~M.} \bibnamefont{Basko}},
  \bibinfo{author}{\bibfnamefont{M.~A.} \bibnamefont{Skvortsov}},
  \bibnamefont{and} \bibinfo{author}{\bibfnamefont{V.~E.}
  \bibnamefont{Kravtsov}}, \bibinfo{journal}{Phys. Rev. Lett.}
  \textbf{\bibinfo{volume}{90}}, \bibinfo{pages}{096801}
  (\bibinfo{year}{2003}).

\bibitem[{\citenamefont{Ringot et~al.}(1999)\citenamefont{Ringot, Lecoq,
  Garreau, and Szriftgiser}}]{AP:DiodeMod:EPJD99}
\bibinfo{author}{\bibfnamefont{J.}~\bibnamefont{Ringot}},
  \bibinfo{author}{\bibfnamefont{Y.}~\bibnamefont{Lecoq}},
  \bibinfo{author}{\bibfnamefont{J.~C.} \bibnamefont{Garreau}},
  \bibnamefont{and}
  \bibinfo{author}{\bibfnamefont{P.}~\bibnamefont{Szriftgiser}},
  \bibinfo{journal}{Eur. Phys. J. D} \textbf{\bibinfo{volume}{7}},
  \bibinfo{pages}{285} (\bibinfo{year}{1999}).

\bibitem[{\citenamefont{Ringot et~al.}(2001)\citenamefont{Ringot, Szriftgiser,
  and Garreau}}]{AP:RamanSpectro:PRA01}
\bibinfo{author}{\bibfnamefont{J.}~\bibnamefont{Ringot}},
  \bibinfo{author}{\bibfnamefont{P.}~\bibnamefont{Szriftgiser}},
  \bibnamefont{and} \bibinfo{author}{\bibfnamefont{J.~C.}
  \bibnamefont{Garreau}}, \bibinfo{journal}{Phys. Rev. A}
  \textbf{\bibinfo{volume}{65}}, \bibinfo{pages}{013403}
  (\bibinfo{year}{2001}).

\bibitem[{\citenamefont{Abrahams et~al.}(1979)\citenamefont{Abrahams, Anderson,
  Licciardello, and Ramakrishnan}}]{Abrahams:PRL79}
\bibinfo{author}{\bibfnamefont{E.}~\bibnamefont{Abrahams}},
  \bibinfo{author}{\bibfnamefont{P.~W.} \bibnamefont{Anderson}},
  \bibinfo{author}{\bibfnamefont{D.~C.} \bibnamefont{Licciardello}},
  \bibnamefont{and} \bibinfo{author}{\bibfnamefont{T.~V.}
  \bibnamefont{Ramakrishnan}}, \bibinfo{journal}{Phys. Rev. Lett.}
  \textbf{\bibinfo{volume}{42}}, \bibinfo{pages}{673} (\bibinfo{year}{1979}).

\bibitem[{\citenamefont{Ohtsuki and Kawarabayashi}(1997)}]{Ohtsuki:JPSJ97}
\bibinfo{author}{\bibfnamefont{T.}~\bibnamefont{Ohtsuki}} \bibnamefont{and}
  \bibinfo{author}{\bibfnamefont{T.}~\bibnamefont{Kawarabayashi}},
  \bibinfo{journal}{J. Phys. Soc. Jpn.} \textbf{\bibinfo{volume}{66}},
  \bibinfo{pages}{314} (\bibinfo{year}{1997}).

\bibitem[{\citenamefont{Hu et~al.}(2008)\citenamefont{Hu, Strybulevych, Page,
  Skipetrov, and {van Tiggelen}}}]{bart:NP08}
\bibinfo{author}{\bibfnamefont{H.}~\bibnamefont{Hu}},
  \bibinfo{author}{\bibfnamefont{A.}~\bibnamefont{Strybulevych}},
  \bibinfo{author}{\bibfnamefont{J.~H.} \bibnamefont{Page}},
  \bibinfo{author}{\bibfnamefont{S.~E.} \bibnamefont{Skipetrov}},
  \bibnamefont{and} \bibinfo{author}{\bibfnamefont{B.~A.} \bibnamefont{{van
  Tiggelen}}}, \bibinfo{journal}{Nature Physics} \textbf{\bibinfo{volume}{4}},
  \bibinfo{pages}{945} (\bibinfo{year}{2008}).

\bibitem[{\citenamefont{Altland}(1993)}]{Atland:PRL93}
\bibinfo{author}{\bibfnamefont{A.}~\bibnamefont{Altland}},
  \bibinfo{journal}{Phys. Rev. Lett.} \textbf{\bibinfo{volume}{71}},
  \bibinfo{pages}{69} (\bibinfo{year}{1993}).

\bibitem[{\citenamefont{Shapiro}(1982)}]{Shapiro:PRB82}
\bibinfo{author}{\bibfnamefont{B.}~\bibnamefont{Shapiro}},
  \bibinfo{journal}{Phys. Rev. B} \textbf{\bibinfo{volume}{25}},
  \bibinfo{pages}{4266} (\bibinfo{year}{1982}).

\bibitem[{\citenamefont{Lemari\'e}(2009)}]{Lemarie:these}
\bibinfo{author}{\bibfnamefont{G.}~\bibnamefont{Lemari\'e}}, Ph.D. thesis,
  \bibinfo{school}{Universit\'e P. et M. Curie}, \bibinfo{address}{Paris}
  (\bibinfo{year}{2009}).

\end{thebibliography}
\end{document}